\documentclass[page-classic]{epl2} 
\usepackage{amssymb,amsmath,graphicx,setspace,color,rotating,subfigure,url,times}
\title{Relaxation dynamics of aftershocks after large volatility shocks in the SSEC index}

\author{Guo-Hua Mu\inst{1,2} \and Wei-Xing Zhou\inst{1,2,3,4}}
\shortauthor{G.-H. Mu \etal}

\institute{
  \inst{1} School of Business, East China University of Science and Technology, Shanghai 200237, China\\
  \inst{2} School of Science, East China University of Science and Technology, Shanghai 200237, China\\
  \inst{3} Center for Econophysics Studies, East China University of Science and Technology, Shanghai 200237, China\\
  \inst{4} Research Center of Systems Engineering, East China University of Science and Technology, Shanghai 200237, China
}


\abstract{The relaxation dynamics of aftershocks after large
volatility shocks are investigated based on two high-frequency data
sets of the Shanghai Stock Exchange Composite (SSEC) index. Compared
with previous relevant work, we have defined main financial shocks
based on large volatilities rather than large crashes. We find that
the occurrence rate of aftershocks with the magnitude exceeding a
given threshold for both daily volatility (constructed using
1-minute data) and minutely volatility (using intra-minute data)
decays as a power law. The power-law relaxation exponent increases
with the volatility threshold and is significantly greater than 1.
Taking financial volatility as the counterpart of seismic activity,
the power-law relaxation in financial volatility deviates remarkably
from the Omori law in Geophysics.}

\begin{document}

\maketitle

\section{Introduction}

Financial markets are complex systems, from which numerous empirical
regularities have been documented in the Econophysics community
\cite{Mantegna-Stanley-2000,Bouchaud-Potters-2000,Sornette-2003,Malevergne-Sornette-2006,Zhou-2007}.
A large proportion of these stylized facts deal with volatility,
which is an important measure of risk of financial assets. The
dynamics of asset volatility exhibit significant similarity in
scaling compared to seismic activities such as the Gutenberg-Richter
law \cite{Gutenberg-Richter-1956-BSSA} and Omori's law
\cite{Omori-1994-JCSIUT}. We note that the Gutenberg-Richter law in
Finance \cite{Kapopoulos-Siokis-2005-EL} has deep connection with
the inverse cubic law in the right tail distribution of volatility
\cite{Liu-Gopikrishnan-Cizeau-Meyer-Peng-Stanley-1999-PRE}. Both
scaling laws concern the dynamic behavior of volatility after stock
market crashes.

Indeed, there are quite a few studies on the dynamic behavior of
volatility after stock market crashes. Sornette and coworkers found
that the implied variance of the Standard and Poor's 500 (S\&P 500)
index after the infamous Black Monday (10/19/1987) decays as a power
law decorated with log-periodic oscillations
\cite{Sornette-Johansen-Bouchaud-1996-JPIF}. Lillo and Mantegna
investigated 1-minute logarithmic changes of the S\&P 500 index
during 100 trading days after the Black Monday and found that the
occurrence of events larger than some threshold exhibits power-law
relaxation for different thresholds \cite{Lillo-Mantegna-2004-PA}.
This Omori law was also found to hold after two other crashes on
10/27/1997 and 08/31/1998 \cite{Lillo-Mantegna-2003-PRE}. There is
more evidence from other stock indexes. Sel\c{c}uk investigated
daily index data from 10 emerging stock markets (Argentina, Brazil,
Hong Kong, Indonesia, Korea, Mexico, Philippines, Singapore, Taiwan
and Turkey) and observed Omori's law after two largest crashes in
each market \cite{Selcuk-2004-PA}. Sel\c{c}uk and Gen\c{c}ay
utilized the 5-minute Dow Jones Industrial Average 30 index (DJIA)
and identified Omori's law after 10/08/1998 and 01/03/2001
\cite{Selcuk-Gencay-2006-PA}.

In a recent paper, Weber {\em{et al.}} extended the procedure
adopted in the aforementioned studies on two aspects
\cite{Weber-Wang-VodenskaChitkushev-Havlin-Stanley-2007-PRE}: (1)
They have smoothed the volatility (absolute of return) a moving
average over an aggregated time scale in order to remove
insignificant fluctuations; and (2) They found that the Omori law
holds not only after big crashes but also after intermediate shocks.
These two issues are of essential importance, which enable us to
discuss the Omori law in alternative manners. On one hand, the
volatility can be defined by the average of absolute returns in a
given time interval rather than the absolute of return over the same
time scale. On the other hand, the main shocks in financial markets
are not necessary to be defined by large crashes. Instead, we can
investigate volatility shocks so that rallies are also included
besides crashes.

Based on these considerations, we shall study in this work the
volatility dynamics of the Shanghai Stock Exchange Composite (SSEC)
index after large volatility shocks. The rest of this paper is
organized as follows. We first present preliminary information about
the composition of data sets, the definition of volatility, and the
mathematical description of Omori's law. We then provide an
objective procedure for the identification of volatility shocks and
investigate the aftershock dynamics of daily and minutely volatility
after large shocks for different thresholds.

%
%

\section{Preliminary information}
\label{s1:PreInfo}

\subsection{The data sets}

We analyze two high-frequency data sets recording the SSEC index
$I(t)$ at two different sampling frequencies. The first data set
contains the close prices at the end of every minute during the
continuous double auction so that its sampling interval is one
minute. This data set covers five and half years from February 2001
to August 2006. The size of the data exceeds 318,602.

The sampling frequency of the second data set fluctuates along time
and is roughly 10 realizations per minute. This data set records
every quotations during the continuous double auction (from 9:30
a.m. to 11:30 a.m. and from 13:00 p.m. to 15:00 p.m.) released by
the Shanghai Stock Exchange and displayed on the terminals that
investors can watch on. The data span from January 2004 to June
2006. The size of the data set is 1,253,440. It is worth noting that
this data set is not ultra-high-frequency but high-frequency, since
ultra-high-frequency data record all transactions
\cite{Engle-2000-Em}.

\subsection{Definition of volatility}

The return $r(t)$ over time scale $\delta t$ is defined as follows
\begin{equation}
 r(t)=\ln[I(t)]-\ln[I(t-\delta t)]~,
 \label{Eq:return}
\end{equation}
where $\delta t$ is time resolution for each data set (one minute
for the first data set and one step in unit of event time for the
second one). In the Econophysics literature including the
aforementioned work on financial Omori law, the absolute of return
$|r(t)|$ is frequently adopted as a measure of volatility on the
time interval $(t-\Delta{t},t]$. Actually, there are various methods
proposed for estimating daily volatility in financial markets
utilizing intraday data
\cite{Pasquini-Serva-1999-EL,Pasquini-Serva-1999-PA,Pasquini-Serva-2000-EPJB,Bollen-Inder-2002-JEF}.
We utilize the following definition for volatility
\cite{Andersen-Bollerslev-Diebold-Labys-2001-JASA,Andersen-Bollerslev-Diebold-Labys-2001-JFE,Bollen-Inder-2002-JEF}
\begin{equation}
 V(t)=\left[\sum_{t-\Delta{t}<\tau\leqslant{t}}{r^2(\tau)}\right]^{1/2}~,
 \label{Eq:Vt}
\end{equation}
which is the root of the sum of squared returns\footnote{The
definition we adopt is different slightly from that of Ref.
\cite{Andersen-Bollerslev-Diebold-Labys-2001-JASA,Andersen-Bollerslev-Diebold-Labys-2001-JFE,Bollen-Inder-2002-JEF},
in which the intraday returns are filtered with an MA(1) model and
the variance $V^2(t)$ is considered instead of $V(t)$.}. In this
work, we consider minutely and daily volatilities. For daily
volatility ($\Delta{t}$ is one day), we adopt the first data set so
that $\delta{t}$ is one minute. For minutely volatility ($\Delta{t}$
is one minute), we use the second data set so that $\delta{t}$
equals to one step of event time.

\subsection{Omori's law vs. power-law relaxation}

It is well-known that there is long memory and clustering in the
volatility. Large shocks in the volatility are often followed by a
series of aftershocks and the occurrence number of events with the
volatility exceeding a given threshold decreases with time.
Recently, the so-called Omori's law was borrowed from Geophysics to
describe the dynamics of financial aftershocks. Omori's law states
that the number of aftershocks decays with some power law of the
time after large shocks: $n(t)\propto t^{-p}$. In order to avoid
divergence at $t=0$, Omori's law is often rewritten as
\begin{equation}
n(t)= K(t+\tau)^{-p}~,
 \label{Eq:nt}
\end{equation}
where $K$ and $\tau$ are two positive constants, and $n(t)$ is the
occurrence rate of aftershocks during $(0,t]$. Equivalently, the
cumulative number of aftershocks after large volatility shocks can
be expressed as follows
\begin{equation}
N(t)= \left\{
 \begin{array}{lll}
 K[(t+\tau)^{1-p}-\tau^{1-p}]/(1-p)~, &~~& p\neq1\\
 K\ln(t/\tau+1)~,&~~&p=1
 \end{array}
 \right..
 \label{Eq:Nt}
\end{equation}
In the empirical analysis, the parameters $K$, $p$ and $\tau$ are
estimated using nonlinear least-squares regression.

\section{Empirical analysis}
\label{s1:EmpAna}

\subsection{Identifying large volatility shocks}

When crashes are concerned, one have to address the problem how to
define a crash, on which a consensus is still lack. A quite feasible
and unambiguous definition is based on large drawdowns
\cite{Johansen-Sornette-1998-EPJB,Johansen-Sornette-2001-JR,Sornette-2003-PR}.
An alternative option is to seek for large price drops within
different time windows \cite{Mishkin-White-2002-NBER}, which was
essentially the same idea used by Sel\c{c}uk \cite{Selcuk-2004-PA}.
A third method is to investigate those large price drops identified
as crashes by academics and professionals. These three methods
identify partially overlapping examples of crashes, that were used
in the previous studies of the dynamics of volatility after crashes
\cite{Lillo-Mantegna-2004-PA,Lillo-Mantegna-2003-PRE,Selcuk-2004-PA,Selcuk-Gencay-2006-PA,Weber-Wang-VodenskaChitkushev-Havlin-Stanley-2007-PRE}.

The situation is different in this study. We concern with large
volatility shocks, which correspond not only to crashes but also to
rallies. For the daily volatility, we first select the seven largest
volatilities. If the time interval between two events are smaller
than 30 trading days, the smaller one is excluded as a foreshock or
aftershock. We then determine the duration of the impact by
identifying the local minimum of volatility after the main shock
within 60 trading days. This selection procedure is detailed in the
following.

\begin{figure}[htb]
\centering
\includegraphics[width=6.5cm]{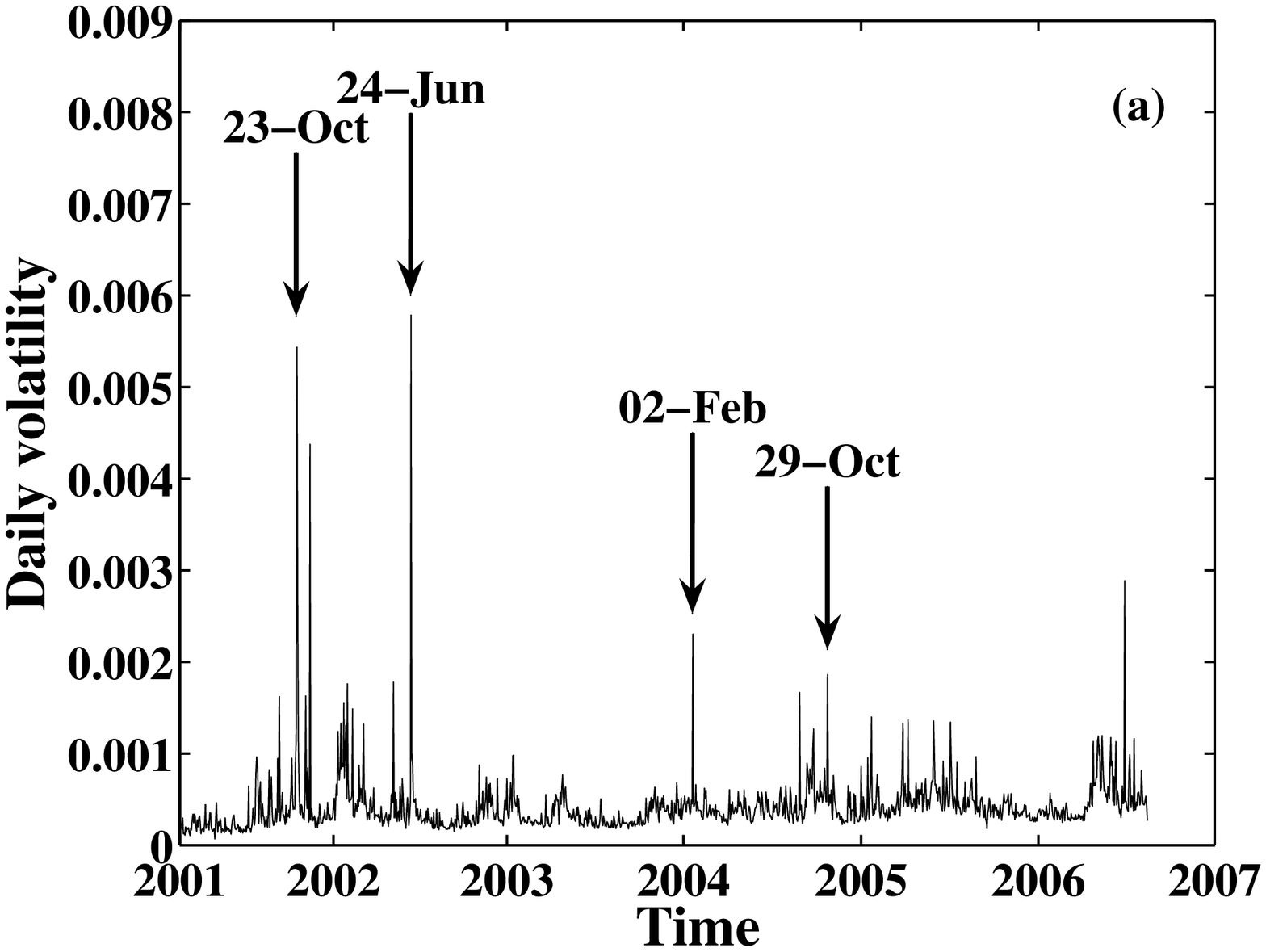}
\includegraphics[width=6.5cm]{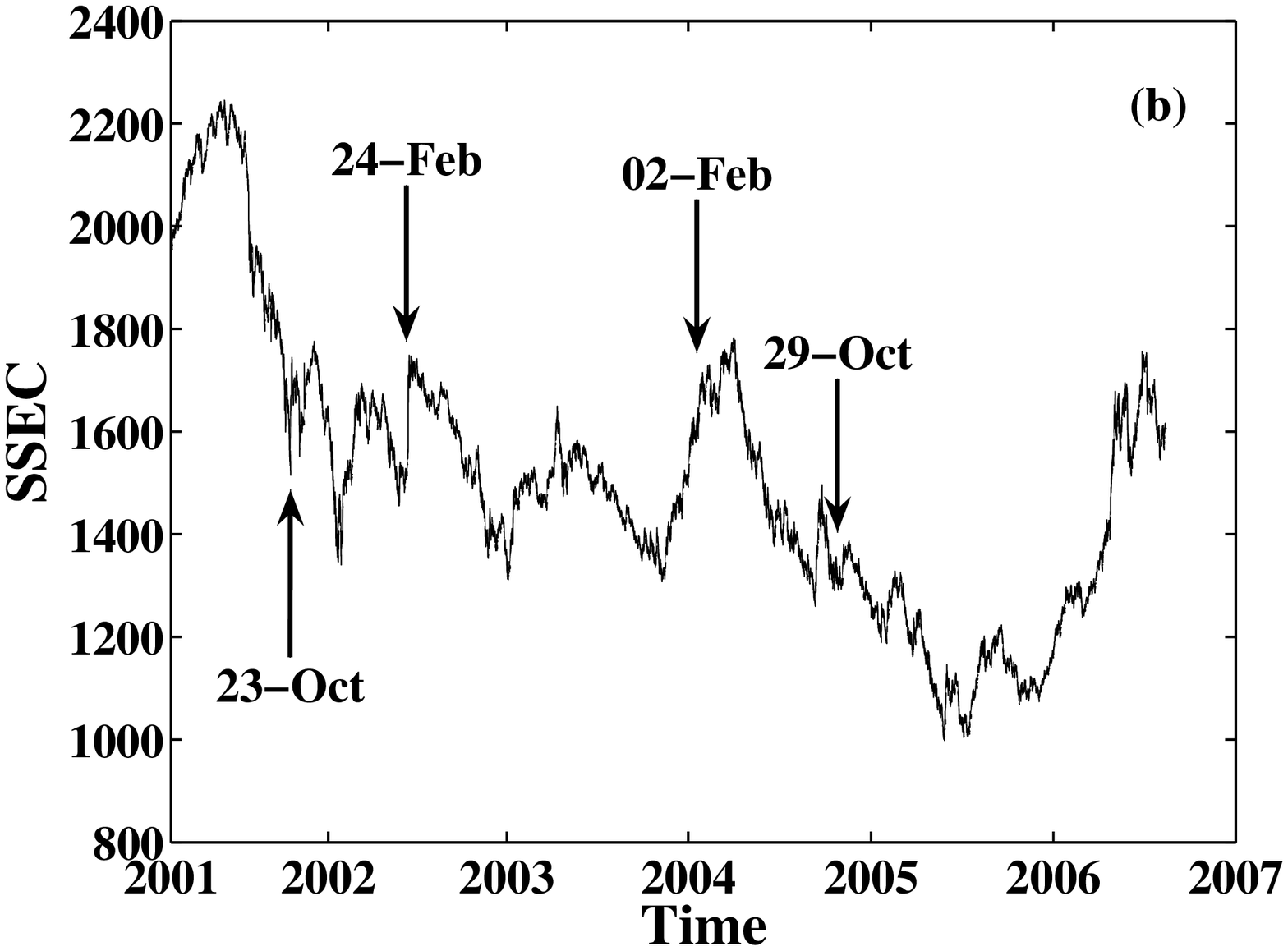}
\caption{\label{Fig:DayV} Identification of four large volatility
shocks. The four dates are indicated in the daily volatility series
(a) and in the evolution of SSEC index (b).}
\end{figure}

Figure \ref{Fig:DayV}(a) illustrates the evolution of daily
volatility for the SSEC index constructed from minutely returns. The
initial seven days are 10/23/2001, 11/16/2001, 05/21/2002,
06/24/2002, 02/02/2004, 10/29/2004 and 07/05/2006, respectively. We
observe that 11/16/2001 is very close to 10/23/2001. Since the
volatility of 10/23/2001 is larger than that of 11/16/2001, we treat
the former as a main shock and the latter its aftershock. Similarly,
the dates 05/21/2002 and 06/24/2002 are also close to each other and
the volatility of the latter is much larger than that of the former.
The latter is thus regarded as the main shock and the former is
considered as its foreshock. Since 06/05/2006 is near the end of the
sample period, it is excluded from investigation. We are thus left
with four events on 10/23/2001, 06/24/2002, 02/02/2004 and
10/29/2004 for analysis, as indicated in Fig.~\ref{Fig:DayV}(a) with
arrows.

Figure \ref{Fig:DayV}(b) shows the locations of the four selected
events in the price trajectory of the SSEC index. It is worthy of
noting that four pieces of information took place on those four
days. The Chinese stock market entered an antibubble regime since
August 2001, which was triggered by the promulgation of the
{\em{Tentative Administrative Measures for Raising Social Security
Funds through the Sale of State-Owned Shares}} by the State Council
of China on 06/24/2001 \cite{Zhou-Sornette-2004a-PA}. China's
Securities Regulatory Commission, however, suspended the fifth rule
of the ``Provisional Rules'' in the evening of 10/22/2001 and the
SSEC index rose up by 9.86\%\footnote{There is a price limit of
$\pm10\%$ fluctuation compared with the closure price on the last
trading day. A rise of 9.86\% in the SSEC means that most of the
stocks went up by 10\%.}. On 2002/06/24, the State Council of China
announced the removal with immediate effect of the provisions of the
{\em{Tentative Administrative Measures for Raising Social Security
Funds through the Sale of State-Owned Shares}}. This caused an
increase of 9.25\% in the SSEC index. In the weekend right before
02/01/2004 (Monday), the State Council of China released {\em{Some
Opinions of the State Council on Promoting the Reform, Opening and
Steady Growth of Capital Markets}} [{\em{Effective}}] and the market
fluctuated remarkably with a 2.08\% rise. In the same year, the
decision of the People's Bank of China to raise the benchmark
lending and deposit interest rates, lift the ceiling of RMB loan
interest rates, and allow a downward movement of RMB deposit
interest rates took effect from October 29. On the same day, the
market swang a lot and closed with a -1.58\% drop.

Volatility shocks with different magnitude may have distinct
durations of impact. In order to investigate the relaxation behavior
of volatility after a shock, it is crucial to determine this impact
duration. The simplest but crude way would be to fix the impact
duration for all shocks, say, 60 or 100 trading days. The
shortcoming of this rule is obvious. In this work, we use a
relatively objective way in which the local minimum of volatility
after an identified shock is regarded as the end of its impact.
Specifically, the local minimum is determined within a
60-trading-day window ensuing the shock. The main information of the
four identified large volatility shocks are presented in Table
\ref{TB:dailyV}. As expected, the impact duration of the main shock
increases approximately with the relative magnitude
$V_{max}/\sigma$, where $\sigma$ is the sample average of the
volatilities. In addition, it is interesting to note that the four
large volatility shocks are more relevant to rallies rather than
crashes.

\begin{table}[htbp]
  \centering
  \caption{Description of the four large volatility shocks for the SSEC index.
      The sample standard deviation of daily volatility $\sigma=0.000421$.
      $t_0$ is the date of volatility shock and $t_1$ is the last date of shock impact.
      $T$ is the duration of shock impact from $t_0$ to $t_1$ in unit of trading day.
      $V_{\max}$ and $V_{\min}$ are the volatilities at time $t_0$ and $t_1$, respectively.
      $r(t_0)$ is the daily return on day $t_0$ referenced to the previous trading day.}
  \label{TB:dailyV}
  \medskip
  \begin{tabular}{ccccccc}
    \hline\hline
    $t_0$ & $t_1$ & $T$ & $V_{max}/\sigma$ & $V_{max}$ & $V_{min}$ & $r(t_0)$ \\
    \hline
    10/23/2001 & 11/29/2001 & 28 & 12.9 & $0.00543$ & $0.00019$ & $+9.86\%$ \\
    06/24/2002 & 09/13/2002 & 60 & 13.7 & $0.00578$ & $0.00015$ & $+9.25\%$\\
    01/02/2004 & 04/07/2004 & 48 &  5.5 & $0.00230$ & $0.00022$ & $+2.08\%$\\
    10/29/2004 & 11/25/2004 & 20 &  4.4 & $0.00186$ & $0.00023$ & $-1.58\%$\\
    \hline\hline
  \end{tabular}
\end{table}

In order to investigate the relaxation behavior of large volatility
shocks at different time scales, we also calculated minutely
volatility and used the same shocks identified in daily volatility
for comparison. Since the minutely time series is shorter, we have
two large shocks left on 01/02/2004 and 10/29/2004. Compared with
the large fluctuations in the Chinese stock market, these two shocks
correspond to neither crashes nor rallies.

\subsection{Aftershock dynamics in daily volatility}

\begin{figure}[htb]
\centering
\includegraphics[width=6cm]{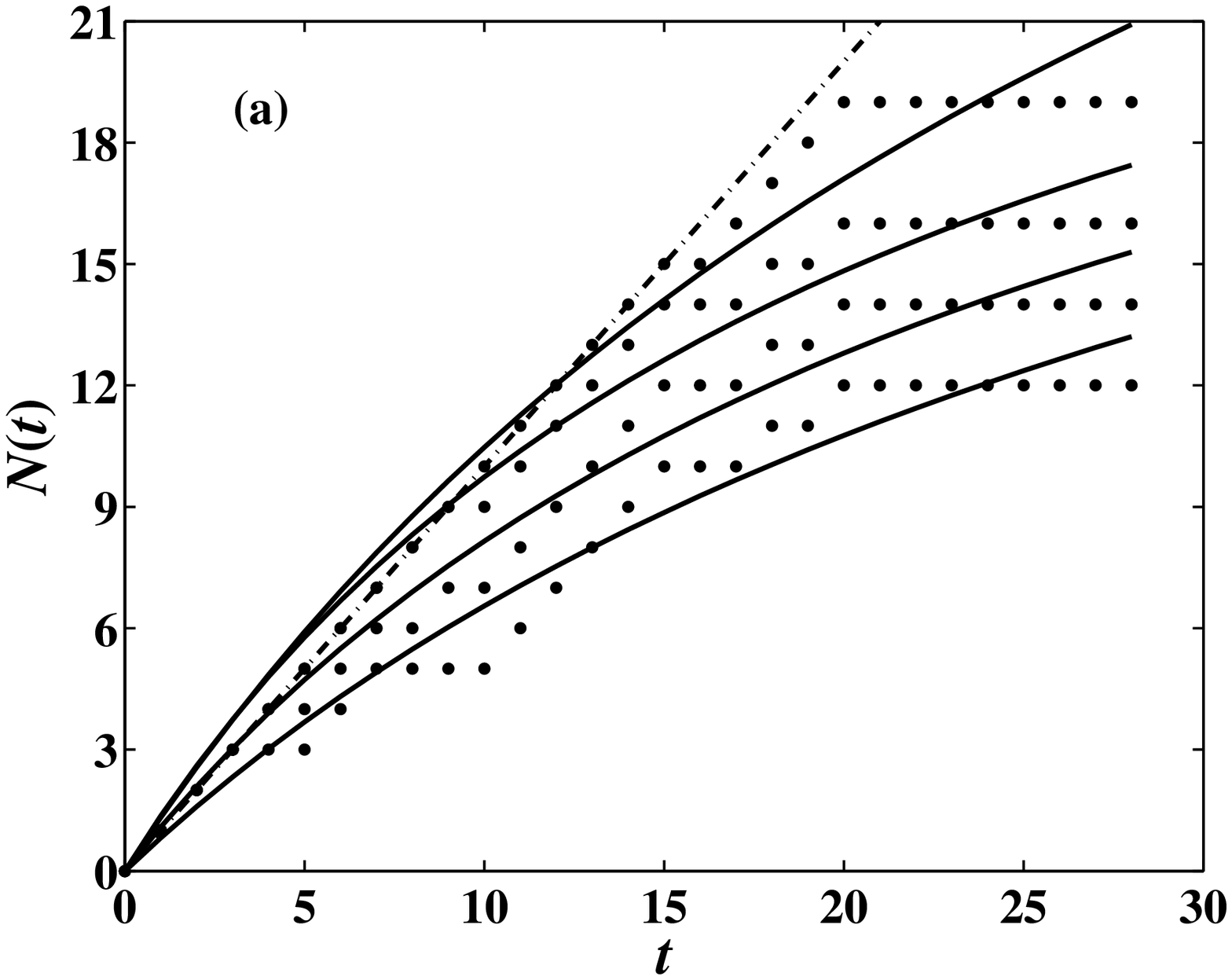}
\includegraphics[width=6cm]{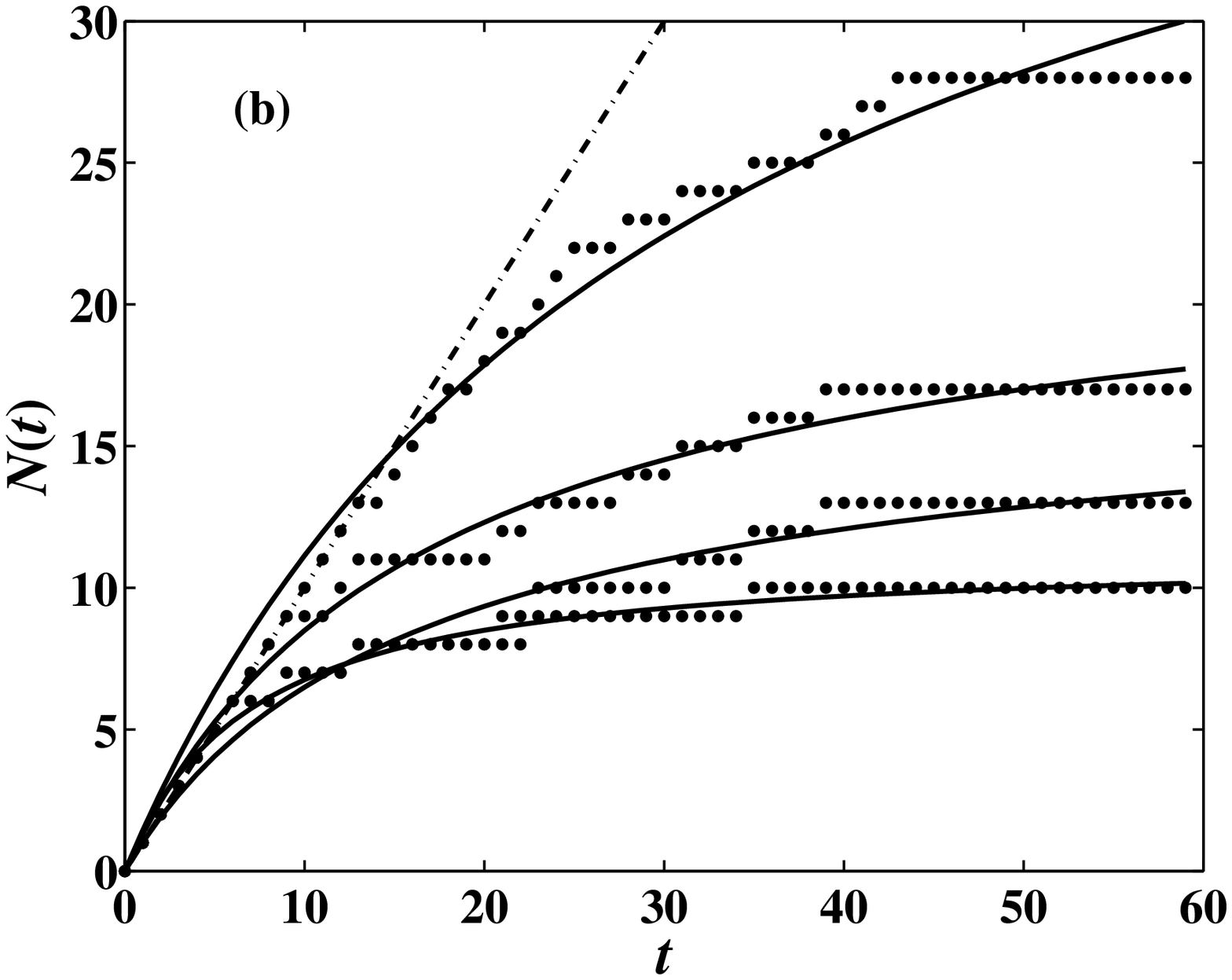}
\includegraphics[width=6cm]{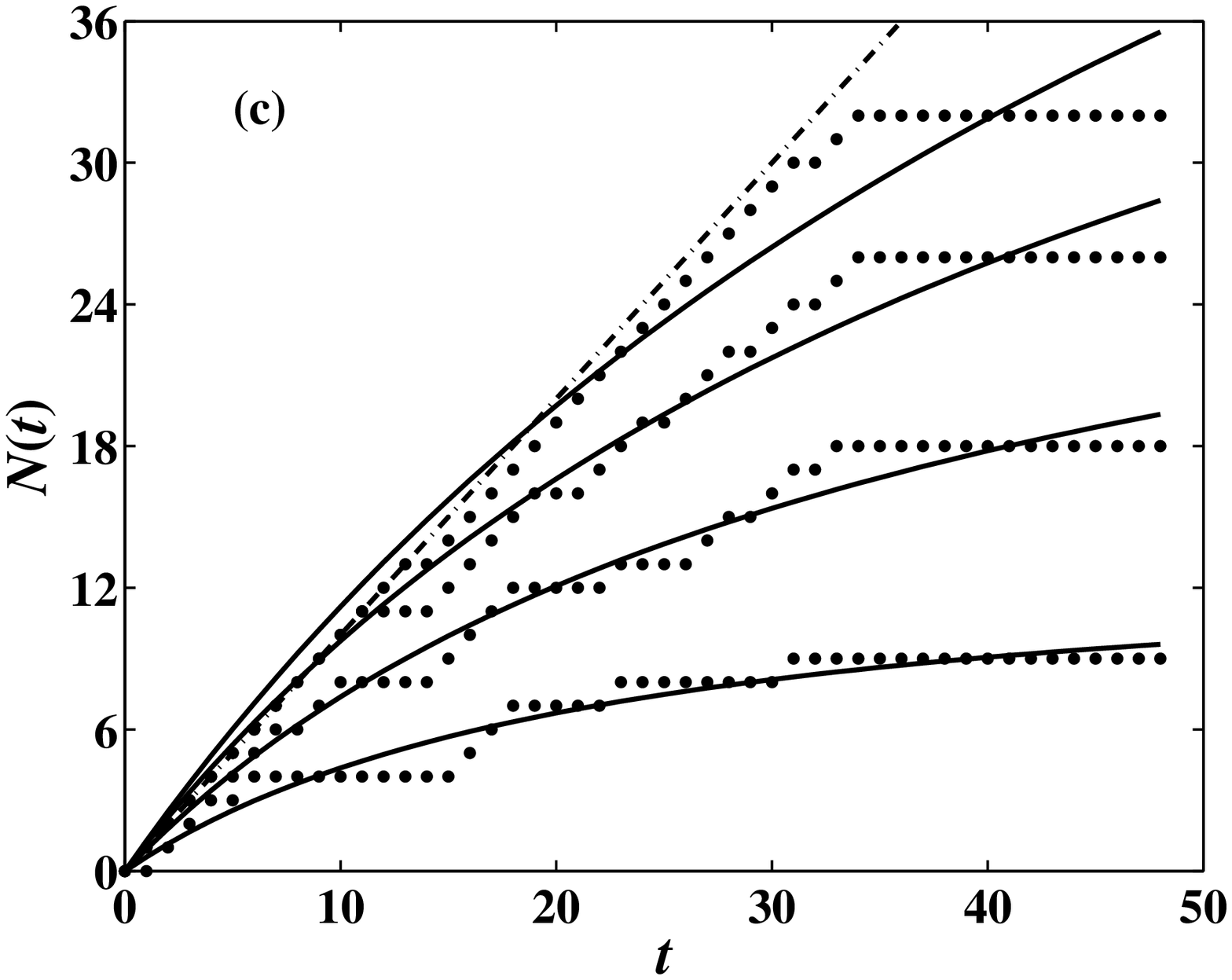}
\includegraphics[width=6cm]{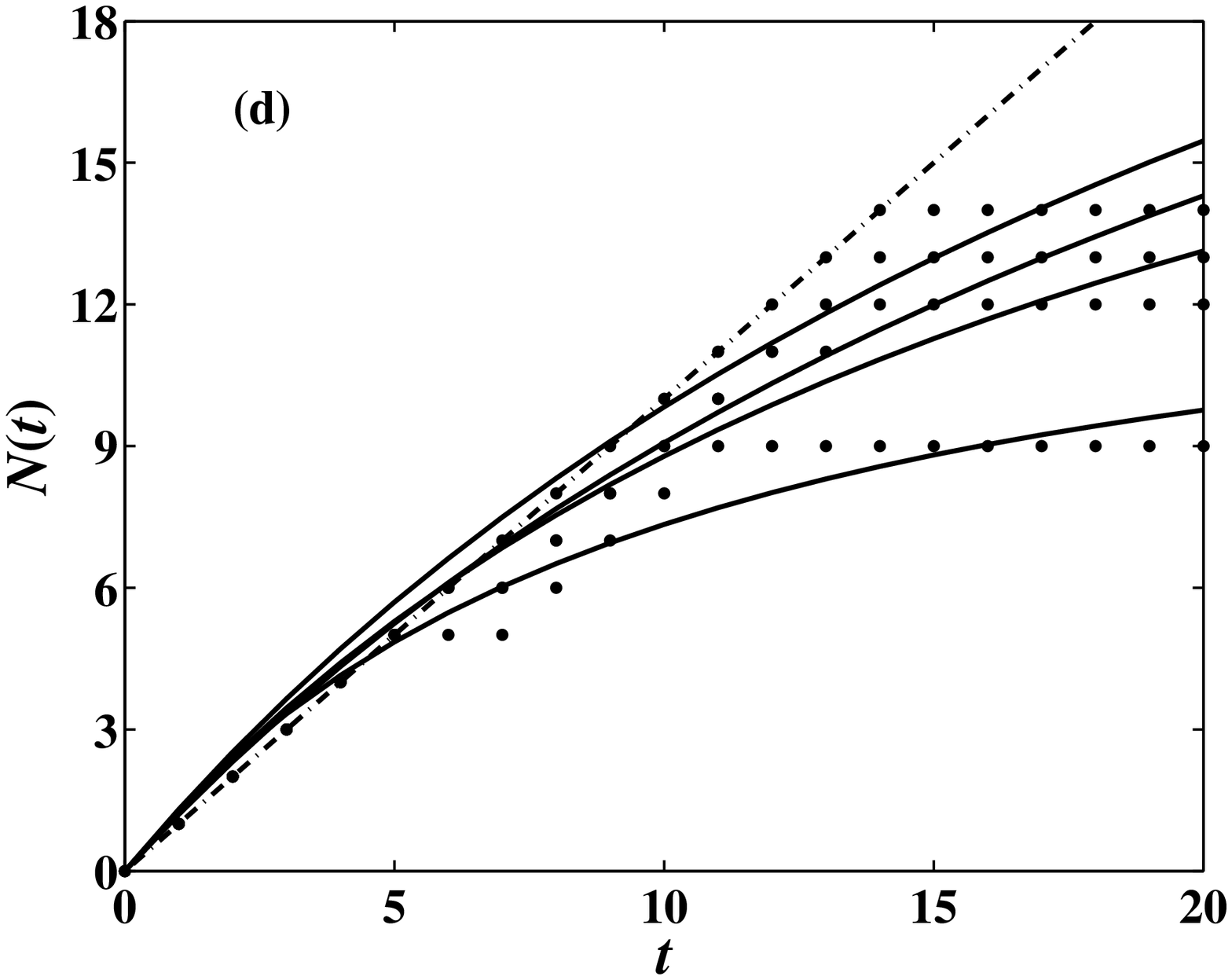}
\caption{\label{Fig:dailyN} Cumulative number $N(t)$ of aftershocks
larger than some fixed threshold after the main shock occurred on
(a) 10/23/2001, (b) 06/24/2002, (c) 02/02/2004, and (d) 10/29/2004.
The sample standard deviation of daily volatility is
$\sigma=4.21\times10^{-4}$. The thresholds are $0.6\sigma$,
$0.7\sigma$, $0.8\sigma$, and $0.9\sigma$ for (a) 10/23/2001, (b)
06/24/2002, and (d) 10/29/2004 and $0.7\sigma$, $0.8\sigma$,
$0.9\sigma$, and $1.0\sigma$ for (c) 02/02/2004. The solid lines are
best fits to the data with Eq.~(\ref{Eq:Nt}). The dashed line is
$N(t)=t$.}
\end{figure}

We have calculated the cumulative number $N(t)$ of aftershocks
larger than some fixed threshold after each main shock. The
threshold $\theta$ is presented based on the sample standard
deviation of daily volatility, which is $\sigma=4.21\times10^{-4}$
within the time period concerned. Four thresholds are selected for
each main shocks: $\theta/\sigma=0.6$, $0.7$, $0.8$, $0.9$ for
10/23/2001, 06/24/2002, and 10/29/2004 and $\theta/\sigma=0.7$,
$0.8$, $0.9$, $1.0$ for 02/02/2004. The resulting functions $N(t)$
are plotted in Fig.~\ref{Fig:dailyN}. Note that the selection of
thresholds is not arbitrary. If $\theta\ll\sigma$, all trading days
are identified as aftershocks such that $N(t)=t$, which is
illustrated in each panel of Fig.~\ref{Fig:dailyN}. If $\theta$ is
too large, $N(t)$ becomes constant shortly after the main shock.
This phenomena can be observed in Fig.~\ref{Fig:dailyN}.

%
%

The power-law relaxation model (\ref{Eq:Nt}) is calibrated for each
empirical $N(t)$ function shown in Fig.~\ref{Fig:dailyN}. The
estimated relaxation exponents $p$ and characteristic time scales
$\tau$ are digested in Table \ref{TB:daily}. All the exponents $p$
are significantly larger than 1 except one case ($t_0=06/24/2002$,
$\theta=0.8\sigma$, $p=0.99$). We notice that these exponents are
much larger than the relaxation exponents of daily volatility for
many other emerging markets \cite{Selcuk-2004-PA}. In other words,
the daily volatility after a main shock relaxes much faster in the
Chinese stock market. In addition, the relaxation exponent $p$
increases with the threshold $\theta$ implying that larger
aftershocks decay faster. This behavior is analogous to that of the
intraday volatility of the S\&P 500 index
\cite{Lillo-Mantegna-2003-PRE}.

\begin{table}[htbp]
  \centering
  \caption{Exponents and characteristic time scales of daily volatility relaxation after
  four large volatility shocks in Chinese stock market.
  The sample mean of daily volatility is $\sigma=4.21\times10^{-4}$.
  The four thresholds for 02/02/2004 are $0.7\sigma$, $0.8\sigma$, $0.9\sigma$, and $1.0\sigma$,
  different from those shown in the table.
  The unit of $\tau$ is trading day.}
  \label{TB:daily}
  \medskip
  \begin{tabular}{ccccccccccccc}
    \hline\hline
      &&  \multicolumn{4}{c}{$p$} & & \multicolumn{4}{c}{$\tau$}       \\%
     \cline{3-6} \cline{8-11}
        $t_0$  && $0.6\sigma$ & $0.7\sigma$ & $0.8\sigma$ & $0.9\sigma$ && $0.6\sigma$ & $0.7\sigma$ & $0.8\sigma$ & $0.9\sigma$
        \\\hline
    10/23/2001 && 1.70 & 1.94 & 1.83 & 2.13 && 28.2 & 20.9 & 23.6 & 38.8\\
    06/24/2002 && 1.86 & 1.86 & 0.99 & 2.14 && 28.1 & 14.8 &  2.1 &  7.9\\
    02/02/2004 && 1.48 & 1.62 & 2.10 & 2.40 && 43.7 & 36.6 & 38.4 & 29.1\\
    10/29/2004 && 1.79 & 1.80 & 1.91 & 2.31 && 23.2 & 23.7 & 18.4 & 12.7\\
    \hline\hline
  \end{tabular}
\end{table}

\subsection{Aftershock dynamics in minutely volatility}

\begin{figure}[htb]
\centering
\includegraphics[width=6.5cm]{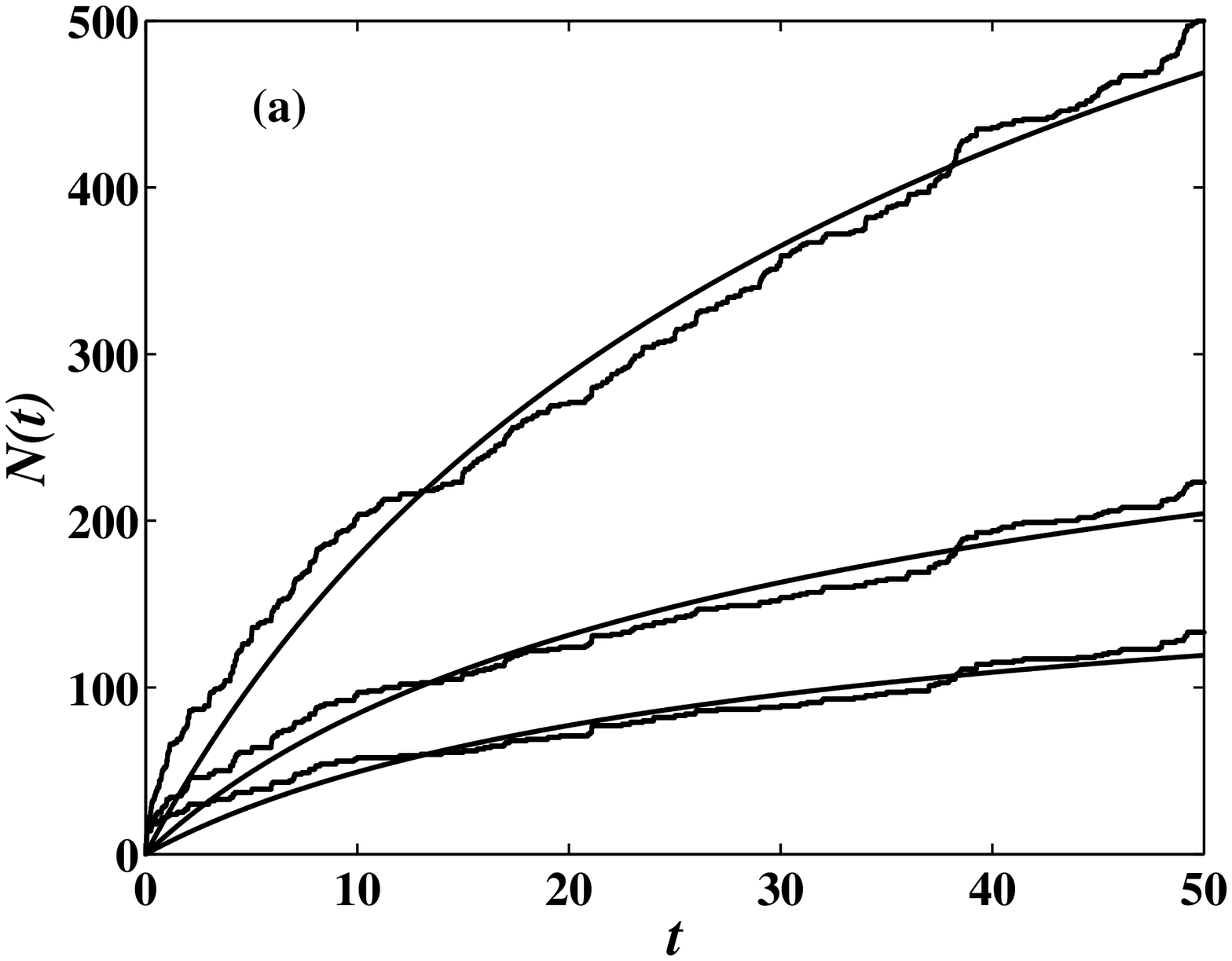}
\includegraphics[width=6.5cm]{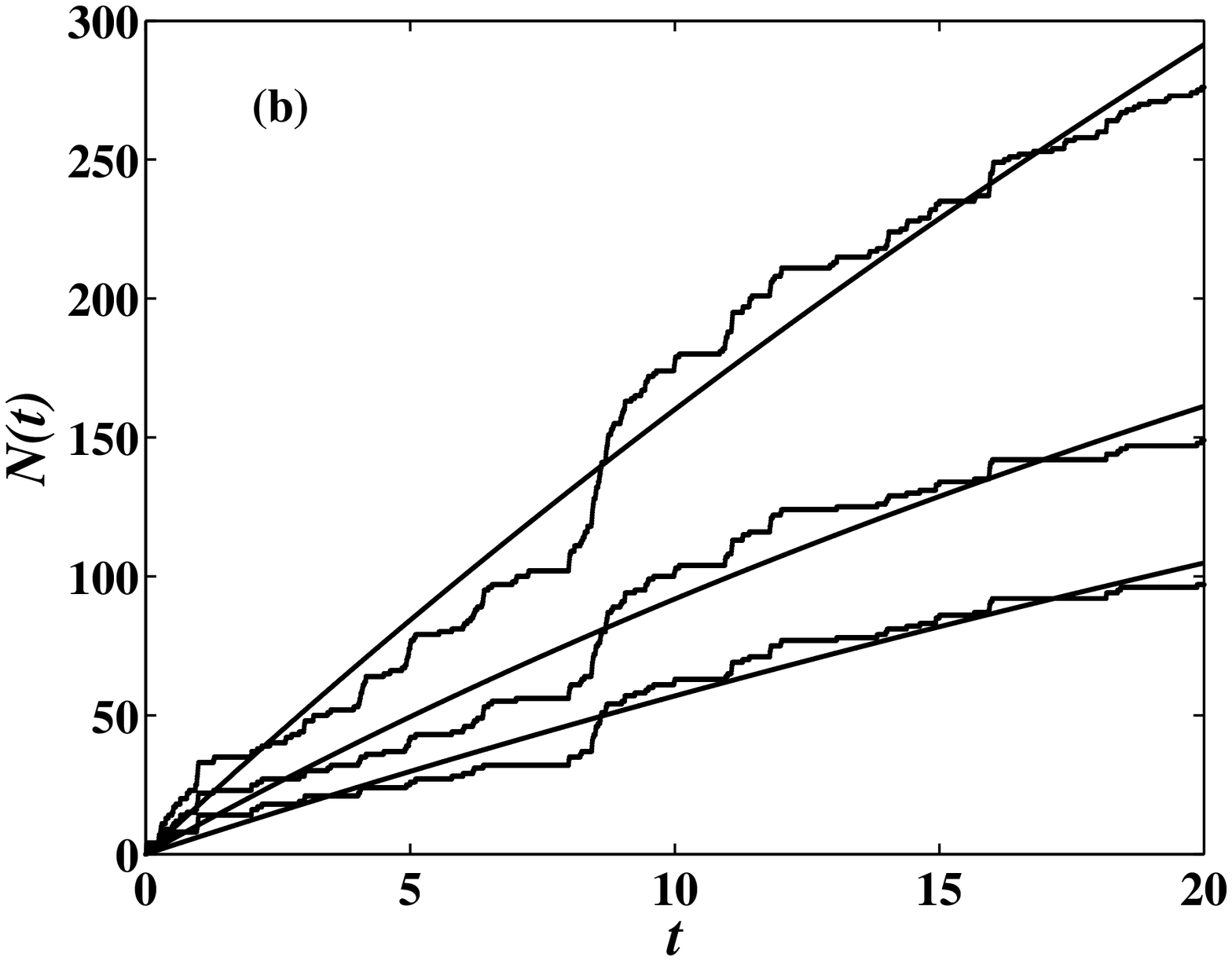}
\caption{\label{Fig:minutelyN} Cumulative number $N(t)$ of
aftershocks larger than some fixed threshold after the main shock
occurred on (a) 02/02/2004 and (b) 10/29/2004. The sample standard
deviation of daily volatility is $\sigma=1.49\times10^{-4}$. The
thresholds are $1.5\sigma$, $1.75\sigma$, and $2.0\sigma$. The solid
lines are best fits to the data with Eq.~(\ref{Eq:Nt}). }
\end{figure}

We have also calculated the cumulative number $N(t)$ of aftershocks
larger than some fixed threshold after each main shock in minutely
SSEC volatility. In this case, the sample standard deviation of
minutely volatility is $\sigma=1.49\times10^{-4}$ within the time
period of the data. Three thresholds are selected for each main
shocks: $\theta/\sigma=1.5$, $1.75$, and $2.0$ for both shocks.
Again, the selection of thresholds is not arbitrary. Smaller
thresholds give straight lines $N(t)\approx{t}$, while too large
thresholds produce constant $N(t)$ shortly after the main shock. The
resulting functions $N(t)$ are plotted in Fig.~\ref{Fig:minutelyN}.
We observe that the three curves for 02/02/2004 on panel (a) of
Fig.~\ref{Fig:minutelyN} exhibit nice power-law curvature. However,
the three curves on panel (b) are not different remarkably from
straight lines. Hence, the minutely SSEC volatility relaxation after
big shocks behaves differently from that for the S\&P 500 index, in
which power-law behavior is unveiled after mediate shocks
\cite{Weber-Wang-VodenskaChitkushev-Havlin-Stanley-2007-PRE}. We
submit that the difference between the construction of volatility in
the two studies can not account for this discrepancy.

We fitted the empirical $N(t)$ functions to Eq.~(\ref{Eq:Nt}), as
shown in Fig.~\ref{Fig:minutelyN} with smooth curves.  The estimated
relaxation exponents $p$ and characteristic time scales $\tau$ are
digested in Table \ref{TB:minutely}. All the exponents $p$ except
for $p=0.97$ are significantly larger than 1. Similar to the daily
volatility case, the relaxation exponent $p$ for minutely volatility
increases with the threshold $\theta$. This observation is different
from those in other stock markets
\cite{Lillo-Mantegna-2004-PA,Lillo-Mantegna-2003-PRE,Selcuk-Gencay-2006-PA,Weber-Wang-VodenskaChitkushev-Havlin-Stanley-2007-PRE}.
It is interesting to note that the characteristic timescale $\tau$
for 10/29/2004 is quite large. Applying Taylor's expansion, we have
$(1+t/\tau)^{1-p}=1+(1-p)(t/\tau)+o(t/\tau)$ and $\ln(t/\tau+1)
=t/\tau+o(t/\tau)$ when $t\ll\tau$. Instituting these two linear
approximations into Eq.~(\ref{Eq:Nt}) gives
\begin{equation}
 N(t) = K\tau^{-p}t
 \label{Eq:Nt2}
\end{equation}
This simple algebraic derivation explains the almost-linear behavior
of $N(t)$ for 10/29/2004 shown in Fig.~\ref{Fig:minutelyN}(b). In
this case, $p$ is a parameter controlling the slope of the linear
$N(t)$. Therefore, the observation that $p$ increases with $\theta$
and the fact that the slope of $N(t)$ decreases with $\theta$ by
definition are consistent with each other.

\begin{table}[htbp]
  \centering
  \caption{Exponents and characteristic time scales of minutely volatility relaxation after
  two large volatility shocks in the Chinese stock market.
  The sample mean of minutely volatility is $\sigma=1.49\times10^{-4}$.
  The unit of $\tau$ is trading day.}
  \label{TB:minutely}
  \medskip
  \begin{tabular}{ccccccccccccc}
    \hline\hline
      &&  \multicolumn{3}{c}{$p$} & & \multicolumn{3}{c}{$\tau$}       \\%
     \cline{3-5} \cline{7-9}
        $t_0$  && $1.5\sigma$ & $1.75\sigma$ & $2.0\sigma$ && $1.5\sigma$ & $1.75\sigma$ & $2.0\sigma$
        \\\hline
    02/02/2004 && 0.97 & 1.58 & 1.72 &&  9.4 &  16.3 & 18.0 \\
    10/29/2004 && 1.54 & 1.74 & 1.76 && 72.7 & 104.0 & 43.7 \\
    \hline\hline
  \end{tabular}
\end{table}

\section{Concluding remarks}
\label{s1:conc}

In summary, we have investigated the dynamic behavior of the Chinese
SSEC volatility after large volatility shocks. Daily and minutely
volatilities are considered, which are calculated based on
high-frequency data at finer scales. We stress that large volatility
shocks are adopted as financial earthquake, rather than the crashes.
These large volatility shocks are selected objectively, together
with the durations of shock impact. In this way, no main shock is
qualified as a crash in our analysis. Instead, two rallies are
selected, which have large volatility. The selection of the
volatility threshold $\theta$ is no arbitrary. Too small or too
large $\theta$ gives $N(t)\sim t$ or $N(t)=const.$, respectively.

We have found that the cumulative number $N(t)$ of aftershocks with
magnitude exceeding a given threshold $\theta$ increases as a
power-law to the time distance $t$ to the main shock. The power-law
exponent $p$ value is a increasing function of the volatility
threshold $\theta$. The aftershock dynamics is very different for
the Chinese SSEC index volatility in the sense that most of the
values of $p$ are significantly larger than 1. Hence, the power-law
relaxation behavior is different from the Omori law where $p$ is
close to 1. This study thus adds a new ingredient to the effort in
searching for idiosyncratic behaviors in the Chinese stock markets
\cite{Zhou-Sornette-2004a-PA,Zhou-Yuan-2005-PA,Jiang-Guo-Zhou-2007-EPJB}.

\acknowledgments

This work was partly supported by the National Natural Science
Foundation of China (Grant No. 70501011), the Fok Ying Tong
Education Foundation (Grant No. 101086), and the Shanghai
Rising-Star Program (No. 06QA14015).

\bibliography{E:/Papers/Auxiliary/Bibliography}

\end{document}